\documentclass[a4paper,12pt]{article}
\usepackage[letterpaper, margin=1in]{geometry}\usepackage{times,longtable}
\usepackage{version}
\usepackage{color}
\usepackage{soul}
\usepackage{url}
\usepackage[pdftex]{graphicx}
\usepackage{subfig}

\usepackage[usenames,dvipsnames]{xcolor}

\usepackage{multirow}
\usepackage{amsmath}

\usepackage{float}
\graphicspath{{../OLD_DRAFT/FIGURE_DEF/},{../OLD_DRAFT/}}

\begin{document}

\title{How pairwise coevolutionary models capture the collective residue variability in proteins}
\author{Matteo Figliuzzi, Pierre Barrat-Charlaix, and Martin Weigt
\footnote{Sorbonne Universit\'es, Universit\'e Pierre-et-Marie-Curie Paris 06, CNRS, Biologie Computationnelle et Quantitative --
Institut de Biologie Paris Seine, 75005 Paris, France}}
\date{}
\maketitle

\begin{abstract}
Global coevolutionary models of homologous protein families, as constructed by direct coupling analysis (DCA), have recently gained popularity in particular due to their capacity to accurately predict residue-residue contacts from sequence information alone, and thereby to facilitate tertiary and quaternary protein structure prediction. More recently, they have also been used to predict fitness effects of amino-acid substitutions in proteins, and to predict evolutionary conserved protein-protein interactions. These models are based on two currently unjustified hypotheses: (a) correlations in the amino-acid usage of different positions are resulting collectively from networks of direct couplings; and (b) pairwise couplings are sufficient to capture the amino-acid variability. Here we propose a highly precise inference scheme based on Boltzmann-machine learning, which allows us to systematically address these hypotheses. We show how correlations are built up in a highly collective way by a large number of coupling paths, which are based on the protein’s three-dimensional structure. We further find that pairwise coevolutionary models capture the collective residue variability across homologous proteins even for quantities which are not imposed by the inference procedure, like three-residue correlations, the clustered structure of protein families in sequence space or the sequence distances between homologs. These findings strongly suggest that pairwise coevolutionary models are actually sufficient to accurately capture the residue variability in homologous protein families.
\end{abstract}

\section{{Introduction}\label{sec:Intro}}

In the course of evolution, proteins may substitute the vast majority of their amino acids without losing their three-dimensional structure and their biological functionality. Rapidly growing sequence databases provide us with ample examples of such evolutionary related, i.e.~homologous proteins, frequently already classified into protein families and aligned into large multiple-sequence alignments (MSA). Typical pairwise sequence identities between homologous proteins go down to 20-30\%, or even below \cite{finn2013pfam}. Such low sequence identities are astonishing since even very few random mutations may destabilize a protein or disrupt its functionality. 
 
Assigning a newly sequenced gene or protein to one of these families helps us to infer functional annotations. Structural homology modeling, e.g., belongs to the most powerful tools for protein-structure prediction \cite{webb2014protein,arnold2006swiss}. However, beyond the transfer of information, the variability of sequences across homologs itself contains information about evolutionary pressures acting in them, and statistical sequence models may unveil that information \cite{durbin1998biological,de2013emerging}.

A first level of information is contained in the variability of individual residues: low variability, i.e. conservation, frequently identifies functionally or structurally important sites in a protein. This information is used by so-called profile models \cite{durbin1998biological}, which reproduce independently the amino-acid statistics in individual MSA columns. They belong to the most successful tools in bioinformatics; they are at the basis of most techniques for multiple-sequence alignment and homology detection, partially as profile Hidden Markov models accounting also for amino-acid insertions and deletions \cite{eddy1998profile}. 

A second level of information is contained in the co-variation between pairs of residues, measurable via the correlated amino-acid usage in pairs of MSA columns \cite{de2013emerging,cocco2017inverse}. Co-variation cannot be captured by profile models, as they treat residues independently. To overcome this limitation, global statistical models with pairwise couplings -- exploiting residue conservation and covariation -- have recently become popular. Methods like the direct coupling analysis (DCA) \cite{weigt2009identification,morcos2011direct}, PsiCov \cite{jones2012psicov} or Gremlin \cite{balakrishnan2011learning} allow for the prediction of residue-residue contacts using sequence information alone, and can be used to predict three-dimensional protein structures \cite{marks2012NATBIOTECH,ovchinnikov2017protein} and to assemble protein complexes \cite{schug2009high,ovchinnikov2014robust,hopf2014elife}. Currently, these methods are the central element of various of the best-performing residue-contact predictors in the CASP competition for protein structure prediction \cite{jones2015metapsicov,wang2017accurate}.
 
Despite their success in practical applications, not much is known about the reasons for this success and their intrinsic limitations. Typically, two hypotheses are made:
{\it (i)} The correlated amino-acid usage in two MSA columns may result from a direct residue-residue contact in the protein structure, causing coordinated amino-acid changes to maintain the protein's stability. It may also result indirectly via intermediate residues, making the direct use of covariation for contact prediction impractical. The success of global models is attributed to their capacity to extract direct couplings from indirect correlations. {\it (ii)} Using the maximum-entropy principle, the simplest models reproducing pairwise residue covariation depend on statistical couplings between residue pairs. Whether or not this model is sufficient to capture also higher-order covariation remains currently unclear. 

So far, these two points have not been investigated systematically. The reason is relatively simple: The inference of pairwise models exactly reproducing the empirical conservation and covariation statistics extracted from an MSA requires to sum over all $20^L$ sequences of aligned length $L$, an unfeasible task for sequences of typical sizes $L=50-500$. Approximation schemes like mean-field approximation \cite{morcos2011direct}, Gaussian approximation \cite{jones2012psicov}, or pseudo-likelihood maximisation \cite{balakrishnan2011learning,ekeberg2013improved}  have been introduced; they perform very well in contact prediction. Their approximate character prohibits, however, the analysis of higher-order correlations and collective effects, since even the pairwise statistics are not well reproduced. More precise approaches have been proposed recently \cite{sutto2015residue,haldane2016structural,barton2016ace}, but their high computational cost has limited applications mostly to anecdotal cases so far.
 
Understanding these basic questions is essential for understanding the success of global coevolutionary models beyond ``black box'' applications, but also for recognising their current limitations and thus potentially to open a way towards improved statistical modeling schemes. To this end, we implement a highly precise approach for parameter inference in pairwise statistical models. Applying this approach to a number of very large protein families (containing sufficient sequences to reliably measure higher-order statistical features), we demonstrate that indirectly generated pair correlations are highly collective effects of entire networks of direct couplings, which are based on the structural vicinity between residues.
 
However, the most interesting finding of the paper is the unexpected accuracy of DCA at reproducing higher-order statistical features, which are not fitted by our approach. These non-fitted features include connected three-point correlations, the distance distributions between natural sequences and between artificial sequences sampled from the model, or the clustered organisation of sequences in sequence space. Currently we do not find indications, that more involved models (e.g.~including three-residue interactions) are needed to reproduce the full sequence statistics: pairwise models are not only necessary as argued above, but seem to be sufficient to describe the sequence variability between homologous proteins. 

\section*{Results}

\subsection*{Direct coupling analysis -- methodology and approximate solutions }

The aim of global coevolutionary sequence models as constructed by DCA is to provide a protein family-specific probability distribution
\begin{equation}
	\label{eq:potts_def}
   P({\underline A}) \ \propto\  \exp\left( \sum_{j>i} J_{ij}(A_i,A_j) + \sum_{i=1}^N h_i(A_i) \right)
\end{equation}
for all full-length amino-acid sequences $\underline{A} = (A_1,\ldots,A_L)$.  To model sequence variability in an MSA, couplings $J_{ij}(A,B)$ and biases (fields) $h_i(A)$ have to be fitted such that model $P(\underline A)$ reproduces the empirically observed frequencies $f_i(A)$ of occurrence of amino acid $A$ in the $i$th MSA column, and co-occurrence $f_{ij}(A,B)$ of amino acids $A$ and $B$ in positions $i$ and $j$ of the same sequence. In other words, the DCA model has to satisfy
\begin{equation}
   \label{eq:maxent_def}
   P_i(A) = f_i(A) \quad \text{and} \quad P_{ij}(A,B) = f_{ij}(A,B)
\end{equation}
for all columns $i,j$ and all amino acids $A,B$, with $P_i$ and $P_{ij}$ being marginal distributions of model $P(\underline A)$, cf. {\it Methods} and Sec.~1 of the {\it Supplement}. 

Eq.~(\ref{eq:maxent_def}) has two important consequences. First, a precisely inferred DCA model reproduces also pairwise connected correlations (or covariances) $c_{ij}(A,B) = f_{ij}(A,B) - f_i(A)f_j(B)$ found in the MSA. This is a crucial difference with profile models, which show vanishing connected correlations by construction. Second, the inference of $P(\underline A)$ via Eq.~(\ref{eq:maxent_def}) does \emph{not} use all the information contained in the MSA, but only the pairwise statistics. For this reason, model $P(\underline A)$ has \emph{a priori} no reason to reproduce any higher order statistics contained in the alignment. In particular, even though a model of the form of Eq.~(\ref{eq:potts_def}) will contain higher-order correlations, such as three-residue correlations, these may differ significantly from those found in the original MSA.

To infer DCA parameters, we need to estimate marginal probabilities for single positions and position pairs from model $P(\underline A)$. Exact calculations of these marginals require to perform exponential sums over $q^L$ terms, with L being the sequence length, and $q=21$ enumerating amino acids and the alignment gap. These sums are infeasible even for short protein sequences, and have been replaced by approximate expressions, e.g. via mean-field \cite{morcos2011direct}, Gaussian \cite{jones2012psicov}, or pseudo-likelihood approximations \cite{balakrishnan2011learning,ekeberg2013improved}. These approximations are sufficiently accurate for residue-contact prediction, which is topological in nature: only the existence of a strong direct statistical coupling has to be detected, not necessarily its precise numerical value. As a consequence, these methods do not reproduce the empirical frequencies and thus do not satisfy Eq.~(\ref{eq:maxent_def}), cf.~Fig.~\ref{fig:fig1} for pseudo-likelihood maximisation (plmDCA \cite{ekeberg2013improved}). More precise methods based on an adaptive cluster expansion \cite{barton2016ace} or Boltzmann machine learning using Markov-chain Monte Carlo sampling \cite{ackley1985learning} for estimating marginal distributions have been proposed recently \cite{sutto2015residue,haldane2016structural}. While decreasing deviations from Eq.~(\ref{eq:maxent_def}) substantially (\emph{i.e.} fitting quality), they are typically much more computationally expensive and not suitable for large-scale application to hundreds or thousands of protein families. 

\subsection*{Accurate fitting is needed to reproduce the empirical residue covariation in homologous protein families}

\begin{figure*}[h!]
\begin{center}
\includegraphics[width=0.95\textwidth]{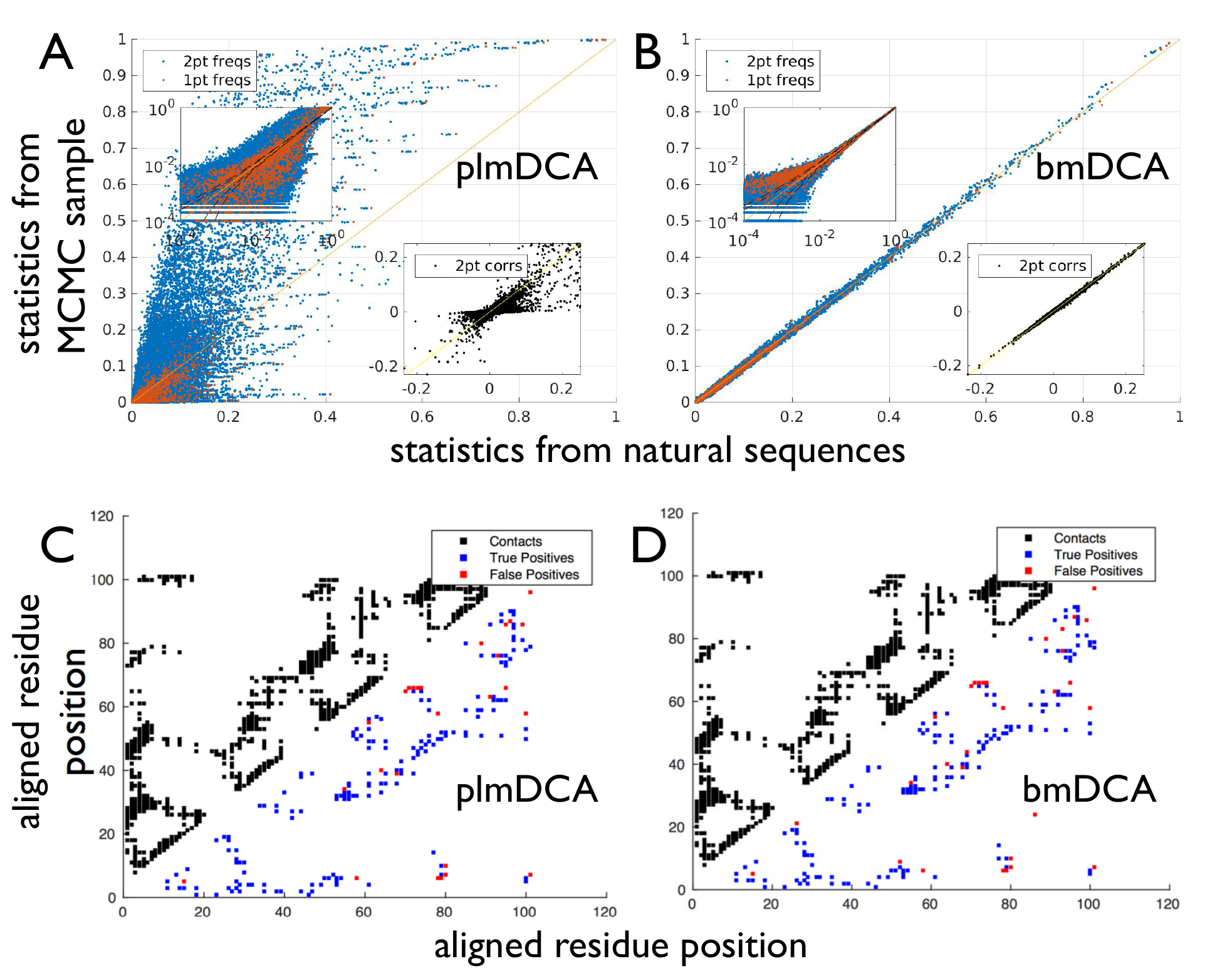}
\end{center}
\caption{{\bf Fitting accuracy and contact prediction for DCA models inferred using pseudo-likelihood maximisation (plmDCA) and Boltzmann-machine learning (bmDCA):} While the Potts model inferred by plmDCA (panel A) fails to reproduce the one- and two-residue frequencies (orange / blue) and the connected two-point correlations (black) in the PF00072 protein family, the model inferred using our bmDCA algorithm (panel B) is very accurate. Slight deviations visible for very small frequencies in log-scale (upper right insert) are results of the $\ell_2$-regularisation penalising strongly negative couplings. Despite these differences, the contact predictions (panel C for plmDCA, panel D for bmDCA) relying on the strongest $2L=224$ DCA couplings (with $|i-j|>4$) are close to identical: native contacts (all-atom distance below 8\AA) are shown in black, predicted contacts in blue (true positives) or red (false positives). Very similar results are observed across all studied protein families, cf.~Sec.~5.1 and 5.5 of the \textit{Supplement}.}
\label{fig:fig1}
\end{figure*}
 
Since the aim of the current paper is to unveil the way DCA disentangles direct couplings and indirect correlations, and to investigate if it captures higher-order statistical observables estimated from the MSA, we have implemented an efficient version of Boltzmann machine (BM) learning described in {\it Methods} and, in full detail, in Sec.~2 of the {\it Supplement}. In short, BM learning estimates the pairwise marginal distributions of $P(\underline A)$ by  Monte-Carlo sampling, and iteratively updates model parameters until Eq.~(\ref{eq:maxent_def}) is satisfied. In contrast to approximations such as applied in plmDCA, the inference of parameters using BM learning can be made arbitrarily accurate, provided that Monte-Carlo samples are large enough and sufficient iterations are performed. In analogy to earlier notation, we will use bmDCA for the resulting implementation of DCA. As is shown in Fig.~\ref{fig:fig1}  and in Table 1, bmDCA reaches very accurate fitting, approaching the statistical uncertainties related to the finite sample size (i.e., the sequence number in each MSA), even for the very large protein families studied here. Obviously bmDCA has a higher computational cost than plmDCA: While plmDCA achieves inference typically in few minutes, bmDCA needs few hours to several days for one family, in dependence of the sequence length and the required fitting accuracy.

Interestingly, the increased fitting accuracy does not improve the contact prediction beyond the one of plmDCA, the currently best unsupervised DCA contact predictor, cf.~Fig.~\ref{fig:fig1}.C+D. Couplings $J_{ij}(A,B)$ are highly correlated between PLM and BM (Pearson correlations of 90\% - 98\% across all studied protein families), in particular large couplings are robust and lead to very similar contact predictions. However, the model statistics depends collectively on all ${\cal O}( q^2 L^2)$ parameters and can thus differ substantially even for small differences in the individual parameters. This sensitivity (so-called criticality) has also been observed in other models inferred from large-scale biological data, cf.~\cite{mora2011biological}.

\begin{table*}[h!]
\begin{center}
\resizebox{\textwidth}{!}{
\begin{tabular}{|c|c|c|c|c|c|c|c|c|c|c|c|c|}
   \hline
   \multicolumn{4}{|c|}{\textbf{protein family}} & \multicolumn{2}{c|}{\textbf{fitting quality}} & \multicolumn{2}{c|}{\textbf{contact prediction}} & \multicolumn{2}{c|}{\textbf{three-point correlations}} & \multicolumn{3}{c|}{\textbf{collectivity of correlations}}  \\
   \hline
   Pfam & $L$ & $M$ & PDB & PLM & BM & PLM & BM & PLM & BM & corr(DI,MI) & corr(L2I,MI) & $\nu$  \\
   \hline
PF00004 & 132 & 39277 & 4D81 & 0.630 & \textbf{0.954} & \textbf{0.672}  & \textbf{0.672} & 0.333  & \textbf{0.980} & 0.33 & 0.42 & 1.2 \\
\hline
PF00005 & 137 & 68891 & 1L7V  & 0.546  & \textbf{0.948} & \textbf{0.599} & 0.586 & 0.718 & \textbf{0.978} &  0.51 & 0.65 & 1.4 \\
   \hline
PF00041 & 85 & 42721 & 3UP1  & 0.897  & \textbf{0.973} & \textbf{0.715} & 0.671 & 0.893 & \textbf{0.991} & 0.61 & 0.77 & 1.7 \\
   \hline
PF00072 & 112 & 73063 & 3ILH  & 0.670  & \textbf{0.978} & 0.836 & \textbf{0.842} & 0.803 & \textbf{0.988} & 0.52 & 0.69 & 1.4\\
   \hline
PF00076 & 69 & 51964 & 2CQD & 0.868  & \textbf{0.977} & \textbf{0.877} & 0.833 & 0.963 & \textbf{0.993} & 0.53 & 0.72 & 1.5\\
   \hline
PF00096 & 23 & 38996 & 2LVH  & 0.954  & \textbf{0.987} & 0.657 & \textbf{0.711} & ND & ND & 0.95 & 0.99  & 2.3 \\
   \hline
PF00153 & 97 & 54582 & 2LCK  & 0.800  & \textbf{0.967} & \textbf{0.601} & 0.563 & 0.517 & \textbf{0.986} & 0.45  & 0.57 & 1.1 \\
   \hline
PF01535 & 31 & 60101 & 4G23  & 0.902  & \textbf{0.994} & 0.630 & \textbf{0.739} & 0.120 & \textbf{0.996} & 0.70 & 0.91 & 1.5\\
   \hline
PF02518 & 111 & 80714 & 3G7E  & 0.624  & \textbf{0.970} & \textbf{0.423} & 0.396 & -0.228 & \textbf{0.986} & 0.47 & 0.60 & 1.6 \\
   \hline
PF07679 & 90 & 36141 & 1FHG & 0.823  & \textbf{0.955} & \textbf{0.826} & \textbf{0.826} & 0.797 & \textbf{0.993} & 0.48 & 0.58 & 1.8\\
   \hline
\end{tabular}
}
\end{center}
\caption{{\bf Results for the 10 selected protein families:}  The first four columns give the ID of the selected protein families together with the sequence length L, alignment depth M and a representative protein structure. The fitting quality measures the Pearson correlation between connected two-point correlations in the natural data, and in a sample drawn from the Potts models inferred by plmDCA and bmDCA (better quality emphasised in boldface). The contact prediction gives the fraction of true positives (all-atom distance $<$ 8\AA) within the first 2L predictions. Columns 9 and 10 provide the Pearson correlation between connected three-point correlations observed in natural and in sampled sequences (due to the dominance of insignificantly small terms, only those with $c_{ijk}^{MSA}(A,B,C)>0.01$ are considered). PF00096, with only 23 aligned positions is the shortest considered protein family, has no significant three-point correlations, neither in the data nor in the Potts model. The last three columns quantify the collective nature of correlations: the Pearson correlation of direct information / mutual information as compared to the length-two information / mutual information, and the exponent of the approximate power-law decay of the strongest paths (in terms of their path information) with their ranking.}
\end{table*}



\subsection*{Indirect correlations result collectively from networks of direct couplings}

bmDCA provides a highly accurate approach to describe the sequence variability of homologous proteins via a pairwise coevolutionary model. This implementation allows us to ask fundamental questions about how DCA works, its capacities and its possible limitations, without being biased by the specificities of approximate DCA implementations.
 
The success of global models as inferred by DCA is typically attributed to the idea that they disentangle statistical correlations, which are empirically observed in an MSA and measured via the mutual information (MI), into a network of direct couplings between residues. The strongest direct couplings are biologically interpretable as residue-residue contacts in the three-dimensional protein structure. However, this idea, even if stated in many papers on the subject, has never been examined in detail, and important questions remain unanswered: can indirect effects be explained by a few strong coupling chains, or are they distributed over networks of numerous small couplings? Are these networks structurally interpretable, i.e.~in relation to a protein’s contact map?
 
\subsubsection*{Correlations are mediated collectively by distributed networks of coevolutionary couplings} 
\begin{figure*}[h!]
\begin{center}
 \includegraphics[width=0.95\textwidth]{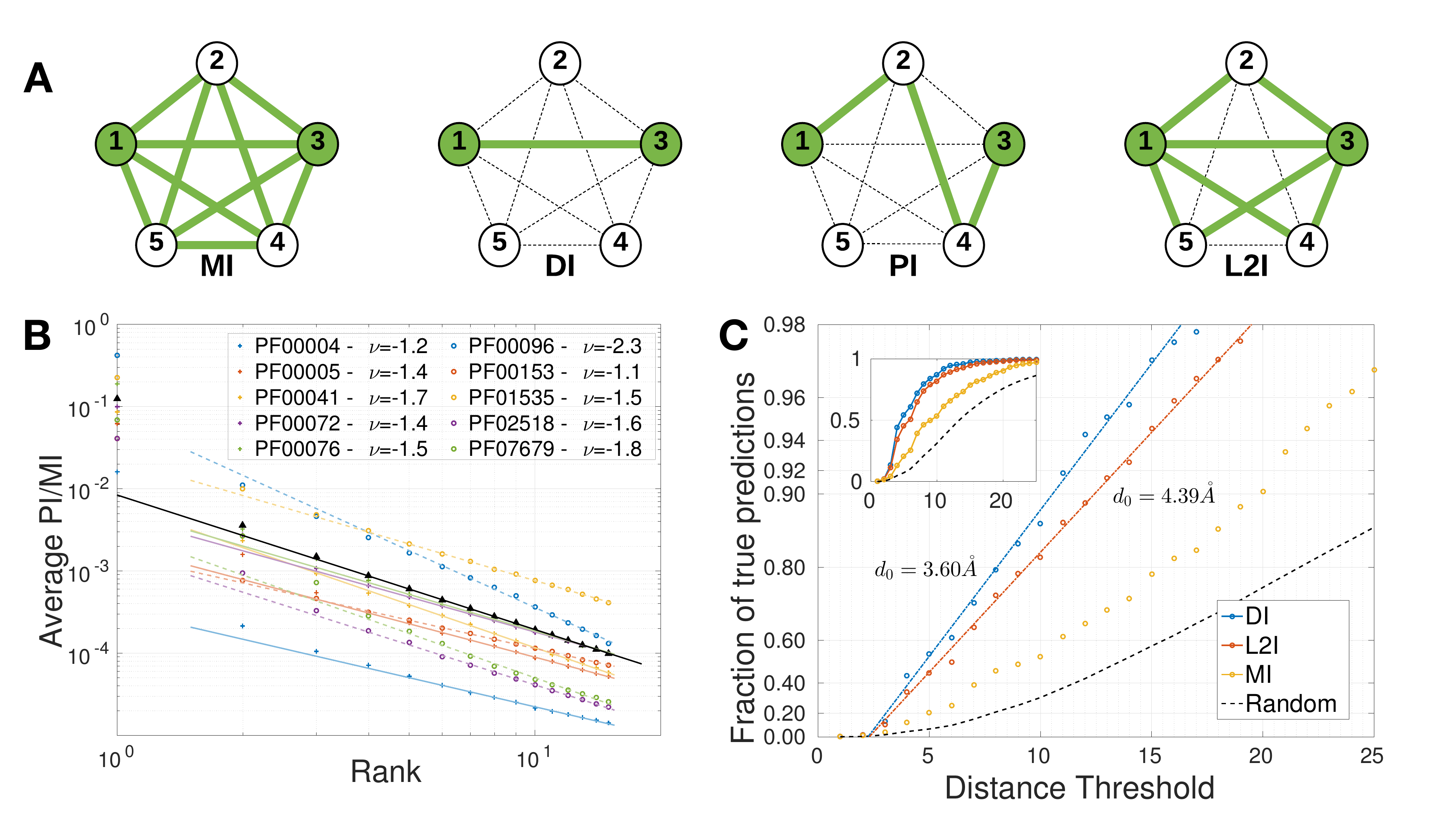}
\end{center}
\caption{{\bf Collective nature of the correlation between two residue positions:} Panel A illustrates the correlation measures used in this work. While the mutual information MI depends collectively on the entire network of coevolutionary couplings, the direct information DI is obtained by taking into account only the single direct coupling between the sites of interest (e.g. 1 and 3 in the figure). All other couplings are formally set to zero. The path information PI is the direct generalisation of DI to the correlation mediated by a single path (e.g. $[1,2,4,3]$ in the figure). The length-two information L2I measures the collective effect of the direct coupling and all length-two paths (e.g. $[1,k,3]$ with $k=2,4,5$). Panel B
shows a log-log plot of the average ratio of path information to mutual information (black curve: all families, coloured curves: individual behavior of families) as a function of the rank of the corresponding path, showing a very slow (approximately power-law) decay. This illustrates the fact that indirect correlations do not depend on a single (or very few) coupling chains, but are distributed over coupling networks. Panel C shows, for the 25 highest ranking residue pairs according to DI, L2I and MI, the fraction of pairs of distance below $d$, as a function of $d$. The scale on the $y$-axis is logarithmic, and chosen in a way that functions of the form $1-e^{-d/d_0}$ will appear as straight lines, the insert shows a standard linear scale. For DI and PI, these curves show a clear exponential convergence to 1, with characteristic distance scales of  3.6 resp. 4.4\AA. MI does not show any exponential behavior, and thus no characteristic distance scale.}
\label{fig:fig2}
\end{figure*}

To answer the first question, we need to quantify the correlation induced by a coupling chain of arbitrary length, connecting any two residues. To this aim, we take inspiration from the concept of direct information (DI) introduced in \cite{weigt2009identification}. DI is a proxy of the strength of the direct interaction $J_{ij}$ between two residue positions $i$ and $j$; it measures the correlation that $i$ and $j$ would have if they were only connected by $J_{ij}$, cf.~Fig.~\ref{fig:fig2}.A. To measure the indirect correlation between $i$ and $j$ induced via a chain of intermediate residues, we introduce the concept of path information (PI), as illustrated again in Fig.~\ref{fig:fig2}.A and defined in {\it Methods}. Now, for each protein family, we extracted the 100 most correlated residue pairs (highest MI). Using a modification of Dijkstra's shortest-path algorithm \cite{dijkstra1959note}  -- which becomes approximate due to the non-additivity of PI but delivers highly reliable results as shown in Sec.~3 of the {\it Supplement} -- we extracted for each residues pairs the 15 strongest coupling paths (highest PI) connecting the two residues. 

In Fig.~\ref{fig:fig2}.B, we show that the decrease of the average strength of the $k$th strongest path is compatible with a slowly decreasing power law, $\langle PI(k) \rangle \propto k^{-\nu}$, with exponents $\nu$  between 1.1 and 2.3.  While this fit is only approximate, as visible by the strong deviations for the strongest path at $k=1$, its slow decay clearly shows that the correlation between two residues typically is not mediated by one or few coupling chains. On the contrary, indirect effects emerge collectively, in the sense that a large number of partially overlapping coupling chains have to be taken into account, each one contributing only a small fraction to the total correlation. It is important to note that the strongest path (rank $k=1$) is on average much stronger than the others and clearly does not fall onto a power law. For the overwhelming majority of the pairs, this strongest path is the direct one containing only one coupling. Its contribution to the total correlation is, on average, about 12,5\% of the total MI. This average is dominated by the shortest protein families, PF00096 and PF01535, who are expected to show less collectivity due to their small number $L$ of aligned residues.


\subsubsection*{On the structural basis of coevolutionary coupling networks} 
As a consequence of the last section, we need to consider the collective effect of multiple paths rather than trying to biologically interpret individual paths beyond the direct one. While this is technically very hard in general, the collective effect of all paths of length two is efficiently computable, cf.~{\it Methods}. The corresponding correlation measure, named here length-two information (L2I) and illustrated in Fig. ~\ref{fig:fig2}.A, adds the $L-2$ possible indirect paths of length 2 (one intermediate residue) to the direct path between two residue positions. As expected, L2I captures a much higher fraction of the full mutual information than DI, cf.~Table 1. However, a large fraction of the mutual information is not yet covered. It is contributed by longer coupling chains: L2I depends only on $2L-3$ out of the $L(L-1)/2$ couplings between residue pairs. Consistent with this observation, the correlation of L2I with MI is much larger in small proteins, and decreases when going to larger proteins.
 
L2I allows for an interesting structural interpretation. It is well established that large DI are good predictors for native residue contacts. Is large L2I a good predictor of second neighbours in the protein structure, i.e.~of residue pairs which are “two contacts away”? To investigate this question, the blue line in Fig.~\ref{fig:fig2}.C displays the fraction of true positive predictions (positive predictive value, averaged over the protein ensemble) within the highest 25 DI as a function of a distance cutoff $d$, which varies between 1\AA\ and 25\AA. It starts at 0 for small $d$, and approaches 1 exponentially with a scaling $1-\exp(-d/d_0)$ of characteristic length $d_0 = 3.6$\AA. At 8\AA\ distance (typically used as contact definition in DCA studies), an accurate prediction of about 85\% true positives (TP) and only 15\% false positives (FP) is reached. Measuring the cut-off dependent positive predictive value for the length-two information L2I, we find again an exponential behavior but with characteristic length $d_0 = 4.4$\AA. The fraction of TP therefore reaches 85\% only between 11 and 12\AA, a distance compatible with second structural neighbours. The finding that the top DI are dominated by direct contacts, and large L2I by residue pairs which are up to second neighbours in the structure, further underlines the structural basis of coevolutionary constraints as captured by DCA. We also note that the full correlation MI -- depending on coupling chains of all possible lengths -- does not imply an exponential behavior in Fig.~\ref{fig:fig2}, and no characteristic length scale can be identified.

\subsection*{Pairwise coevolutionary models accurately reproduce the residue variability beyond the fitted two-residue statistics}
Profile models assuming independent residues are not able to extract the full information contained in the MSA of a protein family. In particular, the inclusion of pairwise coevolutionary couplings is required for the prediction of intra- or inter-protein residue-residue contacts, which has become the most important application of coevolutionary modeling. Furthermore, studies about protein mutational effects \cite{levy2017potts} and the prediction of protein-protein interactions \cite{szurmant2018inter} have underlined the importance of pairwise couplings. 
 
Is there information hidden in large MSA, which cannot be captured by pairwise models? Does one need to include higher-order couplings into the modeling? The highly accurate inference of pairwise models obtained by bmDCA, reproducing faithfully the empirical first- and second-order statistics, allows to address these questions systematically. To this aim, we use MCMC samples from the inferred models to {\it compare statistical observables, which are not a direct consequence of the fitted covariances}. These comparisons unveil the astonishing capacity of bmDCA to capture local and global statistical features, which are not explicitly fitted by the model: pairwise couplings are not only necessary for characterising sequence variability between homologs, but they also seem to be sufficient.
 
\begin{figure*}[h!]
\begin{center}
\includegraphics[width=0.95\textwidth]{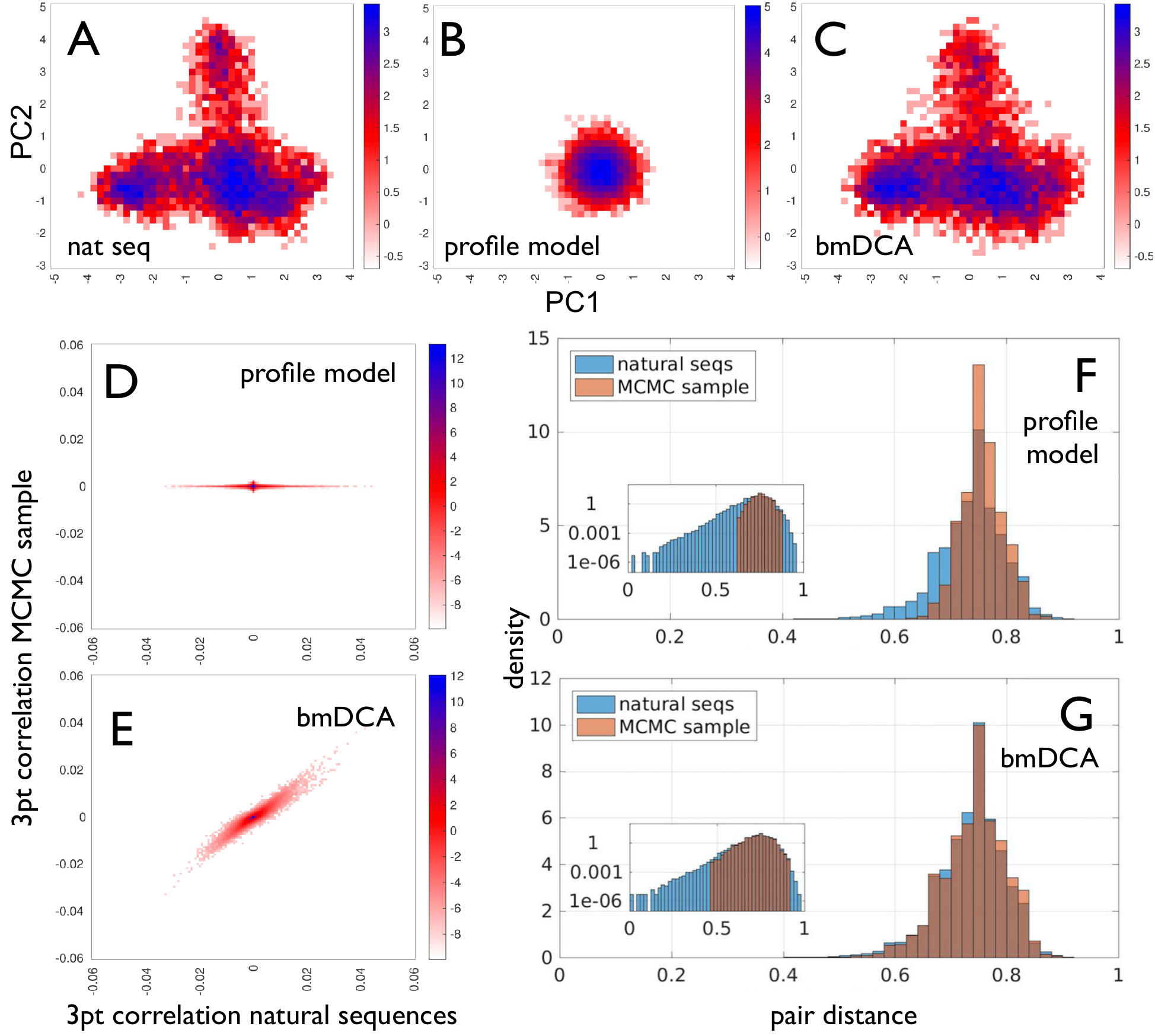}
\end{center}
\caption{{\bf Non-fitted statistical observables are captured by DCA:} Panels A-C: Natural sequences (PF00072 -- A) and MCMC samples from inferred profile (B) and bmDCA(C) models are projected on the first two principal components of the natural MSA. Panels D-E: Three-point correlations of samples of the profile (D) and bmDCA (E) models, as compared to the three-points correlations in the natural sequences. Panels F-G: Histograms of all pairwise Hamming distances between natural or MCMC sampled sequences, for profile (F) and bmDCA (G) models. Surprisingly bmDCA is able to reproduce all three non-fitted statistical properties of the natural MSA, with the difference of the small distances between close homologs, while the profile model not taking into account residue-residue couplings does not. This suggests that accurately inferred pairwise models are necessary and sufficient to capture the residue variability in families of homologous proteins. Similar results are observed across all studied protein families, as is documented in Secs. 5.2-5.4 of the \textit{Supplement}.}
\label{fig:fig3}
\end{figure*}
 
First, we observe that the {\it three-residue statistics} is accurately reproduced by our model including only pairwise couplings: Fig.~\ref{fig:fig3} (cf.~{\it Supplement}, Sec. 5.2, for other families) shows a density-coloured scatter plot of the connected three-point correlations of the natural sequences vs. the MCMC sample drawn from the model. Correlations are high across all protein families for the pairwise model, with close to perfect Pearson correlations ranging from 0.978 to 0.997, cf.~Table 1. Profile models, which by definition do not have any connected three-point correlation, can be seen as null model testing the strength of three-point correlations emerging due to finite sampling. As is shown in Fig.~\ref{fig:fig3}.D, they are at least one order of magnitude smaller than those found empirically, underlining the significance of our findings. The only exception is family PF00096, where no significant connected three-point correlations are detectable in the MSA or in the sample. Note that we use connected correlations $c_{ijk}(A,B,C) = f_{ijk}(A,B,C)-f_{ij}(A,B) f_k(C) - f_{ik}(A,C) f_j(B) - f_{jk}(B,C) f_i(A) +2 f_i(A)f_j(B)f_k(C)$, which are intrinsically harder to reproduce than three-point frequencies $f_{ijk}(A,B,C)$. Note also that our result is far from being obvious: a Gaussian model with the same covariances would have vanishing three-point correlations, while the sequence data and the sample from our DCA model do not. Further more, it is easy to construct models with discrete variables, whose three-point correlations are not reproduced by a pairwise DCA model. This is shown in Sec.~4 of the {\it Supplement}, via analytical calculations and numerical simulations.
 
To complement the three-point statistics, we investigated more global quantities. The first one is the clustered organisation of protein families in sequence space. Fig.~\ref{fig:fig3}.A shows all sequences mapped onto their first two principal components for PF00072 (cf.~{\it Supplement} for other families). We observe a clear clustering into at least three distinct subfamilies, which identify different functional subclasses of the PF00072 protein family (single domain vs. multi-domain architectures with distinct DNA-binding domains). A sample drawn from a profile model does not reproduce this clustered structure (panel B), while the MCMC sample of the bmDCA model does, including the fine structure of the clusters (C). Again, this structure is not a simple consequence of the empirical covariance matrix as a sample from a Gaussian model with the same covariances would not show any clustering.

As a last measure, we compared the pairwise Hamming distances between sequences in the natural MSA and in the model-generated sequences. Again the pairwise bmDCA model is needed to reproduce the bulk of the empirical distribution of pair distances. Interestingly, a difference between the two becomes visible in the small-distance tail of the histograms in Fig.~\ref{fig:fig3}.G: while natural sequences may be close to identical due to a close phylogenetic relation, small sequence distances are never observed in an equilibrium sample of the bmDCA model, i.e., a part of the phylogenetic bias present in the MSA is avoided by the bmDCA model.

\section*{Discussion}

This paper unveils a number of reasons behind the success of global pairwise models in extracting information from the sequence variability of homologous protein sequences. First, we show that residue-residue correlations actually result from the collective variability of many residues, and are not the result of a few strong coupling chains. Therefore local statistical measures taking into account only a small numbers of residues at a time (like correlation measures) are necessarily limited in their capacity to represent the data, and global modeling approaches are needed.
 
One of the most astonishing findings is that many features of the data, which are not explicitly fitted by a pairwise modeling, are nevertheless well reproduced by the inferred models. This includes higher-order correlations, like the connected three-point correlations considered here, and more general aspects of the distribution of amino-acid sequences like the histogram of pairwise Hamming distances between pairs of sequences or the clustered organisation of the sample in sequence space. Interestingly, only the small distances between phylogenetically closely related sequences are not reproduced in a sample drawn from the inferred DCA model. This capacity to reproduce the sequence variability beyond the fitted empirical observables distinguishes the DCA model (fitting one- and two-residue frequencies) from profile models of independent residues (fitting only one-residue frequencies). While the restriction to pairwise models was initially motivated by the limited availability of sequence data -- three-point correlations require to estimate frequencies for $21^3 = 9261$ combinations of amino acids or gaps -- we find that even for large MSA pairwise models seem to be sufficient to capture collective effects beyond residue pairs. 
 
Note that this argument does not rule out the existence of higher-order residue effects in the underlying evolutionary processes shaping the sequence variability in homologous protein families (cf.~\cite{merchan2016sufficiency,schmidt2017three}). However, their statistical signature is not strong enough to be detectable via deviations from the behavior of a pairwise model, even in the large families considered here. Random samples drawn from a DCA model based exclusively on the knowledge of the empirical one- and two-residue statistics appear to be statistically indistinguishable from natural sequences.

This finding is particularly interesting in the context of work made few years ago by the Ranganathan lab \cite{socolich2005evolutionary,russ2005natural}. Using the small WW domain, they applied a number of diverse procedures to scramble MSA of natural sequences to produce artificial sequences. Scrambling MSA columns to maintain residue conservation while destroying residue correlations, lead in all tested cases to non-folding amino-acid sequences. A procedure maintaining also pairwise correlations lead to a substantial fraction of folding and functional proteins. Later on it has been observed that the functional artificial sequences actually have the highest probabilities within pairwise coevolutionary models \cite{balakrishnan2011learning}. These findings open interesting roads to evolution-guided protein design \cite{reynolds2013evolution}.

Note, however, that the finite size of the input MSA requires to use regularised inference, which penalizes large absolute parameter values. It leads to a small bias visible in Fig.~\ref{fig:fig1}.B: small pair frequencies are slightly but systematically overestimated by DCA. This may smoothen the inferred statistical model, cf.~\cite{Otwinowski03062014} for the related case of inferring epistatic fitness landscapes. As a consequence ``bad'' sequences may be given high probabilities in our model. Based on the findings presented in Figs.~\ref{fig:fig1}.B and \ref{fig:fig3}, we expect these effects to be minor. When increasing the regularisation strength beyond parameters used in this work,  the clustered structure of sampled sequences (Fig.~\ref{fig:fig3}.C) disappears gradually. Data in large MSA allow to use small regularisation, thereby simultaneously limiting overfitting of statistical noise and reducing biases in parameter inference. This may be impossible for small MSA, so the ongoing growth of sequence databases is key for the wide applicability of global statistical sequence models.

One potentially important limitation remains: the distribution of sequences in sequence space is not only determined by functional constraints acting on amino-acid sequences, but also by phylogenetic relations between sequences. Natural sequences are, even beyond the very closely related sequences not reproduced by the DCA model, far from being an independent sample of all possible amino-acid sequences. They are correlated due to finite divergence times between homologs, and due to the human selection bias in sequenced species. Any model reproducing the full empirical statistics of the MSA describes therefore a mixture of functional and phylogenetic correlations, while an ideal model would contain the functional ones and discard the phylogenetic ones. How these can be disentangled remains an important open question.

\section*{Methods}

\subsection*{Protein families}
We have selected 10 protein families of known three-dimensional structure which belong to the largest 20 Pfam families \cite{finn2013pfam}, which are not repeat proteins (i.e. they are not just frequent because repeated many times on the same protein), and have an aligned sequence length below 200 amino acids (for computational reasons), cf. Table 1. Sequences with more than 50 alignment gaps are removed. The resulting sequence numbers are reported in Table 1. The main reason to include only large Pfam families is the possibility to accurately estimate three-point correlations. For each triplet of residue positions, there are $21^3 = 9261$ combinations of amino acids or gaps. Non-systematic tests in smaller protein families show that our main findings of the paper translate directly to these families.

\subsection*{Boltzmann machine learning}

DCA infers a Potts model
\begin{equation}
P(A_1,...,A_L) = \frac1{Z} \exp\left\{ \sum_{i<j} J_{ij}(A_i,A_j) + \sum_i h_i(A_i) \right\}
\end{equation}
reproducing the single- and two-residue frequencies found in the input MSA:
\begin{eqnarray}
\sum_{\{A_k|k\neq i\}} P(A_1,...,A_L) &=& f_i(A_i) \nonumber\\
\sum_{\{A_k|k\neq i,j\}} P(A_1,...,A_L) &=& f_{ij}(A_i,A_j) 
\end{eqnarray}
with empirical frequencies $f_i(A_i)$ and $f_{ij}(A_i,A_j)$ defined respectively as the fraction of sequences in the MSA having amino acid $A_i$ (resp. $A_i$ and $A_j$) in column $i$ (resp. in columns $i$ and $j$) (cf.~Sec.~1 of the {\it Supplement} for a precise definition of these frequency counts, including a sequence weighting to reduce phylogenetic biases). For the sake of contact prediction, this inference can be done with efficient approximation schemes like mean-field of pseudo-likelihood maximisation. The objectives of this work -- to understand the collective variability of the residues -- require a more precise inference based on the classical ideas of Boltzmann-machine learning \cite{ackley1985learning}. It consists of an iterative procedure where\\
(i) for a given set of model parameters $\{J_{ij}, h_i\}$, Markov chain Monte Carlo (MCMC) sampling is used to estimate the one- and two-point frequencies of the model;\\
(ii) parameters are adjusted when the estimated model frequencies deviate from the empirical ones.\\
To reduce finite-sample effects, the model parameters are subject to an $\ell_2$-regularisation. The likelihood function is convex, guaranteeing convergence to a single globally optimal solution, which reproduces the empirical one- and two-point frequencies with arbitrary accuracy. The direct implementation of Boltzmann-machine learning is computationally very slow. We have therefore introduced a reparameterization of the model, which allows to replace the gradient ascent of the likelihood by a faster pseudo-Newtonian method. Technical details of the implementation are described in Sec.~2 of the {\it Supplement}.

\subsection*{From direct couplings to indirect correlations}

\subsubsection*{Quantifying the strength of a coupling chain}

To quantify the strength of a coupling chain, we generalize the direct information introduced in \cite{weigt2009identification}. There, the direct probability 
\begin{equation}
	\label{eq:Pdir_def}
	\begin{split}
	&P^{dir}_{ij}(A_i, A_j) = \\ 
							&\exp\{ J_{ij}(A_i,A_j) + \tilde{h}_i(A_i) + \tilde{h}_j(A_j) \}/Z_{ij}\, .
	\end{split}
\end{equation}
was defined as the hypothetical distribution of two residues $i$ and $j$ connected only by the inferred direct coupling $J_{ij}$ and having the empirical single-residue frequencies $f_i(A_i)$ and $f_j(A_j)$, thereby removing all indirect effects from model $P$. Parameters $\tilde{h}_i$ and $\tilde{h}_j$ are to be adjusted to ensure correct marginals. The \emph{path probability} between positions $i_1$ and $i_{L+1}$ through the length-$L$ path $[i_1, i_2\ldots i_{L+1}]$ is a direct generalisation:
	\begin{equation}
		\label{eq:Ppath_def}
		\begin{split}
		&P^{path}_{[i_1\dots i_{L+1}]}(A_{i_1},A_{i_{L+1}}) = \\ 
		&\sum_{\{A_{i_2}\ldots A_{i_L}\}} f_{i_1}(A_{i_1}) \prod_{l=1}^L P^{dir}_{i_{l+1}i_l}(A_{i_{l+1}} \vert A_{i_l})\, ,
		\end{split}
	\end{equation}
with $P^{dir}_{ij}(A_i \vert A_j) = P^{dir}_{ij}(A_i, A_j) / f_j(A_j)$. Eq.~\ref{eq:Ppath_def} contains the product of direct probabilities for all links in the path, in analogy to a Markov chain. The sum over all configurations taken by intermediate sites $[i_2 \ldots i_L]$ is performed efficiently by dynamic programming; the definition guarantees the empirical marginals in all sites on the path.

To measure the correlation mediated by direct links or indirect paths, we use variants of the mutual information based on the direct and path probabilities. To this aim we define the \emph{direct information} (DI) as 
\begin{equation} \label{eq:DI_def}
\begin{split}
DI_{ij} &=\\
  &\sum_{A_{i},A_{j} =1}^q P^{dir}_{ij}(A_{i},A_{j}) \log \frac{P^{dir}_{ij}(A_{i},A_{j})}{f_i(A_{i})f_j(A_{j})}\, ,
\end{split}
\end{equation}
and the \emph{path information} (PI) as
\begin{equation} \label{eq:PI_def}
\begin{split}
PI_{[i \ldots j]} &=\\
  &\sum_{A_{i},A_{j} =1}^q P^{path}_{[i \ldots j]}(A_{i},A_{j}) \log \frac{P^{path}_{[i \ldots j]}(A_{i},A_{j})}{f_i(A_{i})f_j(A_{j})} \, .
\end{split}
\end{equation}
The full \emph{mutual information} (MI) is defined by replacing $P^{dir}$ or $P^{path}$ by $f_{ij}$.

 \subsubsection*{The joint effect of paths of length 2}

Quantifying the strength of a group of indirect effects between two sites $i$ and $j$ is in general non trivial. However, it is possible if one only considers all chains of couplings that go through at most one intermediary site $k$. In other words, one can combine the direct path $[i j]$ and all the chains of the form $[i k j]$ ($k\neq i,j$) into a single probability distribution: 
 	\begin{equation}
 		\label{eq:Length2_def}
 		P^{L2}_{ij}(A_i,A_j) \propto \frac{P^{dir}(A_i,A_j)}{z_i(A_i) z_j(A_j)}\cdot\prod_{k\neq i,j} P^{path}_{[i\,k\,j]}(A_i,A_j),
 	\end{equation}
where $z_i$ and $z_j$ ensure $P^{L2}_{ij}$ to have marginals $f_i$ and $f_j$. The path probabilities can be simply multiplied since each intermediate residue $k$ appears only once, and they become conditionally independent for given $(A_i,A_j)$.
The correlation resulting from this combination of paths is the mutual information of $P^{L2}_{ij}$, called $L2I$.

\section{Supplementary Material}
Supplementary text and figures are available  at Molecular Biology and Evolution
online (http://www.mbe.oxfordjournals.org/). Code and raw data can be accessed via Github (https://github.com/matteofigliuzzi/bmDCA).

\section{Acknowledgments}

MW acknowledges funding by the ANR project COEVSTAT (ANR-13-BS04-0012-01), and by the European Union's H2020 research and innovation programme MSCA-RISE-2016 under grant agreement No. 734439 INFERNET. This work was undertaken partially in the framework of CalSimLab, supported by the grant ANR-11-LABX-0037-01 as part of the "Investissements d'Avenir" program (ANR-11-IDEX-0004-02).

\bibliographystyle{unsrt}
\bibliography{bib_dca}

\begin{thebibliography}{10}

\bibitem{finn2013pfam}
Robert~D. Finn, Alex Bateman, Jody Clements, Penelope Coggill, Ruth~Y.
  Eberhardt, Sean~R. Eddy, Andreas Heger, Kirstie Hetherington, Liisa Holm,
  Jaina Mistry, Erik L.~L. Sonnhammer, John Tate, and Marco Punta.
\newblock Pfam: the protein families database.
\newblock {\em Nucleic Acids Research}, 42(D1):D222--D230, 2014.

\bibitem{webb2014protein}
Benjamin Webb and Andrej Sali.
\newblock Protein structure modeling with modeller.
\newblock {\em Protein Structure Prediction}, pages 1--15, 2014.

\bibitem{arnold2006swiss}
Konstantin Arnold, Lorenza Bordoli, J{\"u}rgen Kopp, and Torsten Schwede.
\newblock The swiss-model workspace: a web-based environment for protein
  structure homology modelling.
\newblock {\em Bioinformatics}, 22(2):195--201, 2006.

\bibitem{durbin1998biological}
Richard Durbin, Sean~R Eddy, Anders Krogh, and Graeme Mitchison.
\newblock {\em Biological sequence analysis: probabilistic models of proteins
  and nucleic acids}.
\newblock Cambridge university press, 1998.

\bibitem{de2013emerging}
David de~Juan, Florencio Pazos, and Alfonso Valencia.
\newblock Emerging methods in protein co-evolution.
\newblock {\em Nature Reviews Genetics}, 14(4):249--261, 2013.

\bibitem{eddy1998profile}
Sean~R. Eddy.
\newblock Profile hidden markov models.
\newblock {\em Bioinformatics}, 14(9):755--763, 1998.

\bibitem{cocco2017inverse}
Simona Cocco, Christoph Feinauer, Matteo Figliuzzi, Remi Monasson, and Martin
  Weigt.
\newblock Inverse statistical physics of protein sequences: A key issues
  review.
\newblock {\em Report on Progress in Physics}, 2017.

\bibitem{weigt2009identification}
Martin Weigt, Robert~A White, Hendrik Szurmant, James~A Hoch, and Terence Hwa.
\newblock Identification of direct residue contacts in protein--protein
  interaction by message passing.
\newblock {\em Proceedings of the National Academy of Sciences}, 106(1):67--72,
  2009.

\bibitem{morcos2011direct}
Faruck Morcos, Andrea Pagnani, Bryan Lunt, Arianna Bertolino, Debora~S Marks,
  Chris Sander, Riccardo Zecchina, Jos{\'e}~N Onuchic, Terence Hwa, and Martin
  Weigt.
\newblock Direct-coupling analysis of residue coevolution captures native
  contacts across many protein families.
\newblock {\em Proceedings of the National Academy of Sciences},
  108(49):E1293--E1301, 2011.

\bibitem{jones2012psicov}
David~T Jones, Daniel~WA Buchan, Domenico Cozzetto, and Massimiliano Pontil.
\newblock Psicov: precise structural contact prediction using sparse inverse
  covariance estimation on large multiple sequence alignments.
\newblock {\em Bioinformatics}, 28(2):184--190, 2012.

\bibitem{balakrishnan2011learning}
Sivaraman Balakrishnan, Hetunandan Kamisetty, Jaime~G Carbonell, Su-In Lee, and
  Christopher~James Langmead.
\newblock Learning generative models for protein fold families.
\newblock {\em Proteins: Structure, Function, and Bioinformatics},
  79(4):1061--1078, 2011.

\bibitem{marks2012NATBIOTECH}
Debora~S Marks, Thomas~A Hopf, and Chris Sander.
\newblock Protein structure prediction from sequence variation.
\newblock {\em Nature Biotechnology}, 30(11):1072--1080, 2012.

\bibitem{ovchinnikov2017protein}
Sergey Ovchinnikov, Hahnbeom Park, Neha Varghese, Po-Ssu Huang, Georgios~A
  Pavlopoulos, David~E Kim, Hetunandan Kamisetty, Nikos~C Kyrpides, and David
  Baker.
\newblock Protein structure determination using metagenome sequence data.
\newblock {\em Science}, 355(6322):294--298, 2017.

\bibitem{schug2009high}
Alexander Schug, Martin Weigt, Jos{\'e}~N Onuchic, Terence Hwa, and Hendrik
  Szurmant.
\newblock High-resolution protein complexes from integrating genomic
  information with molecular simulation.
\newblock {\em Proceedings of the National Academy of Sciences},
  106(52):22124--22129, 2009.

\bibitem{ovchinnikov2014robust}
Sergey Ovchinnikov, Hetunandan Kamisetty, and David Baker.
\newblock Robust and accurate prediction of residue--residue interactions
  across protein interfaces using evolutionary information.
\newblock {\em Elife}, 3:e02030, 2014.

\bibitem{hopf2014elife}
Thomas~A Hopf, Charlotta~PI Sch{\"a}rfe, Jo{\~a}o~PGLM Rodrigues, Anna~G Green,
  Oliver Kohlbacher, Chris Sander, Alexandre~MJJ Bonvin, and Debora~S Marks.
\newblock Sequence co-evolution gives 3d contacts and structures of protein
  complexes.
\newblock {\em Elife}, 3:e03430, 2014.

\bibitem{jones2015metapsicov}
David~T Jones, Tanya Singh, Tomasz Kosciolek, and Stuart Tetchner.
\newblock Metapsicov: combining coevolution methods for accurate prediction of
  contacts and long range hydrogen bonding in proteins.
\newblock {\em Bioinformatics}, 31(7):999--1006, 2015.

\bibitem{wang2017accurate}
Sheng Wang, Siqi Sun, Zhen Li, Renyu Zhang, and Jinbo Xu.
\newblock Accurate de novo prediction of protein contact map by ultra-deep
  learning model.
\newblock {\em PLOS Computational Biology}, 13(1):e1005324, 2017.

\bibitem{ekeberg2013improved}
Magnus Ekeberg, Cecilia L{\"o}vkvist, Yueheng Lan, Martin Weigt, and Erik
  Aurell.
\newblock Improved contact prediction in proteins: using pseudolikelihoods to
  infer potts models.
\newblock {\em Physical Review E}, 87(1):012707, 2013.

\bibitem{sutto2015residue}
Ludovico Sutto, Simone Marsili, Alfonso Valencia, and Francesco~Luigi Gervasio.
\newblock From residue coevolution to protein conformational ensembles and
  functional dynamics.
\newblock {\em Proceedings of the National Academy of Sciences},
  112(44):13567--13572, 2015.

\bibitem{haldane2016structural}
Allan Haldane, William~F Flynn, Peng He, RSK Vijayan, and Ronald~M Levy.
\newblock Structural propensities of kinase family proteins from a potts model
  of residue co-variation.
\newblock {\em Protein Science}, 25(8):1378--1384, 2016.

\bibitem{barton2016ace}
John~P Barton, Eleonora De~Leonardis, Alice Coucke, and Simona Cocco.
\newblock Ace: adaptive cluster expansion for maximum entropy graphical model
  inference.
\newblock {\em Bioinformatics}, 32(20):3089--3097, 2016.

\bibitem{ackley1985learning}
David~H Ackley, Geoffrey~E Hinton, and Terrence~J Sejnowski.
\newblock A learning algorithm for boltzmann machines.
\newblock {\em Cognitive Science}, 9(1):147--169, 1985.

\bibitem{mora2011biological}
Thierry Mora and William Bialek.
\newblock Are biological systems poised at criticality?
\newblock {\em Journal of Statistical Physics}, 144(2):268--302, 2011.

\bibitem{dijkstra1959note}
Edsger~W Dijkstra.
\newblock A note on two problems in connexion with graphs.
\newblock {\em Numerische mathematik}, 1(1):269--271, 1959.

\bibitem{levy2017potts}
Ronald~M Levy, Allan Haldane, and William~F Flynn.
\newblock Potts hamiltonian models of protein co-variation, free energy
  landscapes, and evolutionary fitness.
\newblock {\em Current Opinion in Structural Biology}, 43:55--62, 2017.

\bibitem{szurmant2018inter}
Hendrik Szurmant and Martin Weigt.
\newblock Inter-residue, inter-protein and inter-family coevolution: bridging
  the scales.
\newblock {\em Current opinion in structural biology}, 50:26--32, 2018.

\bibitem{merchan2016sufficiency}
Lina Merchan and Ilya Nemenman.
\newblock On the sufficiency of pairwise interactions in maximum entropy models
  of networks.
\newblock {\em Journal of Statistical Physics}, 162(5):1294--1308, 2016.

\bibitem{schmidt2017three}
Michael Schmidt and Kay Hamacher.
\newblock Three-body interactions improve contact prediction within
  direct-coupling analysis.
\newblock {\em Physical Review E}, 96(5):052405, 2017.

\bibitem{socolich2005evolutionary}
Michael Socolich, Steve~W Lockless, William~P Russ, Heather Lee, Kevin~H
  Gardner, and Rama Ranganathan.
\newblock Evolutionary information for specifying a protein fold.
\newblock {\em Nature}, 437(7058):512--518, 2005.

\bibitem{russ2005natural}
William~P Russ, Drew~M Lowery, Prashant Mishra, Michael~B Yaffe, and Rama
  Ranganathan.
\newblock Natural-like function in artificial ww domains.
\newblock {\em Nature}, 437(7058):579--583, 2005.

\bibitem{reynolds2013evolution}
Kimberly~A Reynolds, William~P Russ, Michael Socolich, and Rama Ranganathan.
\newblock Evolution-based design of proteins.
\newblock {\em Methods Enzymol}, 523:213--235, 2013.

\bibitem{Otwinowski03062014}
Jakub Otwinowski and Joshua~B. Plotkin.
\newblock Inferring fitness landscapes by regression produces biased estimates
  of epistasis.
\newblock {\em Proceedings of the National Academy of Sciences},
  111(22):E2301--E2309, 2014.

\end{thebibliography}

\end{document}